\documentclass[twocolumn,superscriptaddress,showkeys]{revtex4}
\usepackage{amssymb,epsf,epsfig,amsmath,color, array}
\usepackage{hyperref}
\usepackage{mciteplus}
\usepackage{ifthen}

\newcommand{\lsim}{
\mathrel{\hbox{\rlap{\hbox{\lower4pt\hbox{$\sim$}}}\hbox{$<$}}}}
\newcommand{\gsim}{
\mathrel{\hbox{\rlap{\hbox{\lower4pt\hbox{$\sim$}}}\hbox{$>$}}}}

\usepackage{accents}
\newcommand*{\fancybar}{\scalebox{.4}{(}\raisebox{-1.7pt}{-}\scalebox{.4}{)}}
\newcommand*{\brabar}[1]{\accentset{\fancybar}{#1}}
\newcommand{\rdpi}{\ensuremath{r_{B}^{D\pi}}\,}
\newcommand{\ddpi}{\ensuremath{\delta_{B}^{D\pi}}}
\newcommand{\rdk}{\ensuremath{r_{B}^{DK}}\,}

\def\D0{D\O }

\begin{document}

\setcounter{page}{0}

\date{\today}

\title{\boldmath  Estimating \rdpi as input to the determination of CKM angle $\gamma$\unboldmath}

\author{Matthew Kenzie}
\affiliation{CERN, Geneva, Switzerland}

\author{Maurizio Martinelli}
\affiliation{EPFL, Lausanne, Switzerland}

\author{Niels Tuning}
\affiliation{Nikhef, Amsterdam, The Netherlands}

\begin{abstract}
\vspace{0.2cm}\noindent
  The interference between Cabibbo-favoured and Cabibbo-suppressed $B\to D\pi$ decay amplitudes
  provides sensitivity to the CKM angle $\gamma$.
  The relative size of the interfering amplitudes is an important ingredient in the determination of $\gamma$.
  Using branching fractions from various $B\to Dh$ decays,
  and the measured value for $\rdk$,
  the magnitude of the amplitude ratio of $B^+\to D^0\pi^+$ and  $B^+\to \bar{D}^0\pi^+$ decays
  is estimated to be $\rdpi = 0.0053 \pm 0.0007$.
\end{abstract}


\maketitle

\clearpage
\section{Introduction}
\label{sec:Introduction}

\begin{figure}[!b]
  \begin{center}
    \begin{picture}(250,200)(0,0)
      \put(  0,110){\includegraphics[scale=0.8]{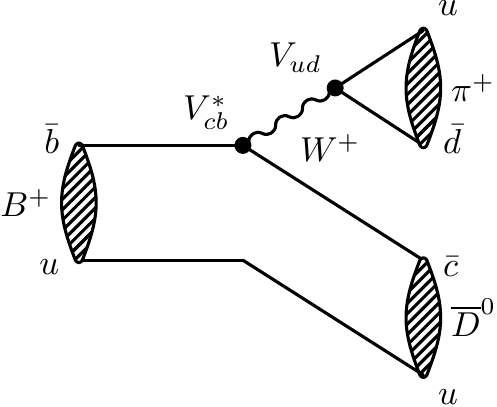}}
      \put(125,110){\includegraphics[scale=0.8]{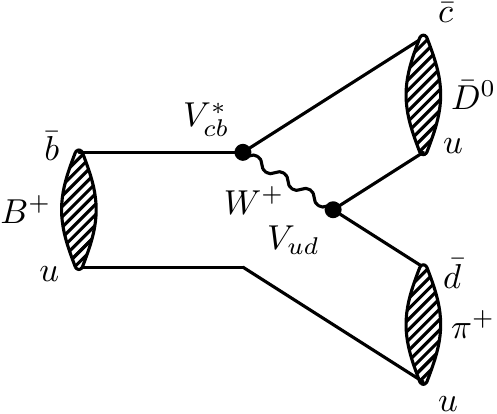}}
      \put(10,210){(a)  $T:$}
      \put(125,210){(b)  $C:$}
      \put(  0,0){\includegraphics[scale=0.8]{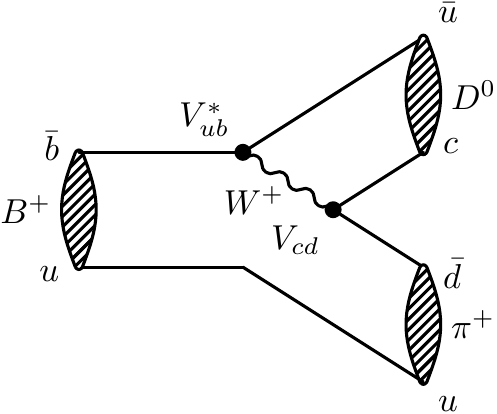}}
      \put(125,0){\includegraphics[scale=0.8]{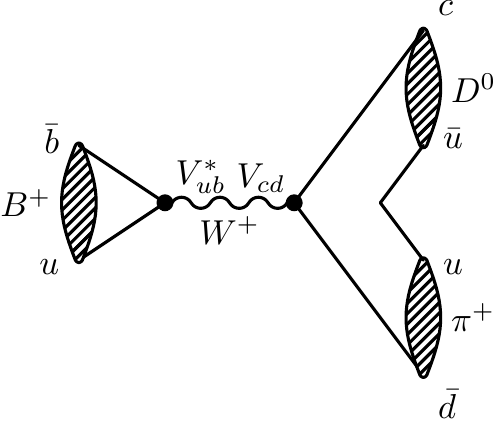}}
      \put(10, 85){(c)  $C^{ub}:$}
      \put(125,85){(d)  $A:$}
    \end{picture}
    \caption[TC]{\em (a,b) The color-allowed (tree) ($T$) and color-suppressed ($C$)
      topologies contributing to $B^+\to \bar{D}^0\pi^+$ amplitude proceeding through $V_{cb}$,
      and (c,d) the color-suppressed ($C^{ub}$) and annihilation topologies contributing to $B^+\to D^0\pi^+$
      amplitude proceeding through $V_{ub}$.}
    \label{fig:top}
  \end{center}
\end{figure}
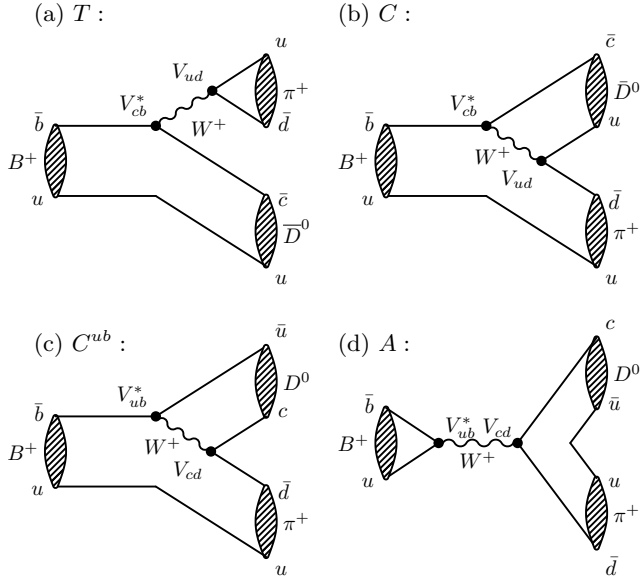

The CKM description of charged-current quark transitions has been experimentally scrutinized
to an impressive accuracy. The CKM angle $\gamma$ encapsulates the relative phase
between $b\to c$ and $b\to u$ quark transitions,  $\gamma \equiv
\arg\left[-\frac{{V_{ud}}\phantom{^*}{V_{ub}}^*}{{V_{cd}}\phantom{^*}{V_{cb}}^*}\right]$, and is
determined with a precision of $7^o$, as compared to a precision below
$3^o$ deduced from indirect measurements~\cite{Charles:2015gya,Bona:2009cj}.

The interference between $B^+\to \bar{D}^0\pi^+$ and $B^+\to {D}^0\pi^+$ decays,
with the ${D}^0$ and $\bar{D}^0$ meson decaying to the same final state (charge-conjugation
is implied throughout), provides
sensitivity to the relative weak phase $\gamma$~\cite{delAmoSanchez:2010dz, Belle:2011ac,LHCB-PAPER-2013-020}.
Experimental determination of $\gamma$ from $B\to D\pi$ like decays
is influenced by the effect of the unknown hadronic parameters: \rdpi, the relative
magnitude of the Cabibbo-suppressed $B^+\to {D}^0\pi^+$ amplitude compared to
the Cabibbo-favored $B^+\to \bar{D}^0\pi^+$ amplitude, and \ddpi, the strong phase
difference between the favored and suppressed modes. The ratio of amplitudes, \rdpi,
determines the size of the interference effect, and hence the sensitivity to
CKM angle $\gamma$.

A previous simultaneous determination of $\gamma$, \rdpi and \ddpi\, from the LHCb
collaboration, using $B\to D\pi$ like modes, found multiple solutions for
\rdpi~\cite{LHCb-CONF-2014-004}. Consequently, an estimate of its magnitude can provide 
useful information to improve the determination of $\gamma$~\cite{LHCb-PAPER-2016-032}.

In this paper we estimate the ratio of amplitudes
$$\rdpi \equiv A(B^+\to {D}^0\pi^+)/A(B^+\to \bar{D}^0\pi^+),$$
using branching fractions from various
$B\to Dh$ decays, that proceed through similar decay topologies~\cite{Fleischer:2011},
and using the measured value of \rdk~\cite{LHCb-PAPER-2016-032}.
A similar approach was used to estimate the ratio of amplitudes for the decays 
$B^0\to D^\pm\pi^\mp$~\cite{DeBruyn:2012jp}.
An overview of the decays used is given in Table~\ref{tab:decays}.
The amplitudes of the decays that involve a kaon in the final state are denoted by primed symbols.

\begin{table}[!b]
\begin{center}
\begin{tabular}{lccr}
\hline
Decay                       & Topology         &BR ($\times 10^{-4}$)~\cite{PDG2014}& CKM factor \\
\hline
\phantom{\Large{B}}\hspace{-0.45cm}
$A(B^+\to \bar{D}^0\pi^+)$  & $T+C$              & $48.1 \pm 1.5   $                   & $V_{cb}V_{ud}$\\
$A(B^0\to \bar{D}^0\pi^0)$  & $(C-E)/\sqrt{2}$ & $2.63 \pm 0.14  $                   & $V_{cb}V_{ud}$\\
\hline
\phantom{\Large{B}}\hspace{-0.45cm}
$A(B^+\to \bar{D}^0K^+)$    & $T'+C'$            & $3.70 \pm 0.17  $                   & $V_{cb}V_{us}$\\
$A(B^0\to \bar{D}^0 K^0)$   & $C'$               & $0.52 \pm 0.07  $                   & $V_{cb}V_{us}$\\
\hline
\phantom{\Large{B}}\hspace{-0.45cm}
$A(B^+\to D_s^+ \phi)$      & $A'$               & $0.017^{+0.012}_{-0.007}  $         & $V_{ub}V_{cs}$\\
\hline
\end{tabular}
\caption{The decays under study are listed, with the topologies contributing to the
amplitude, the branching fraction, and the relevant CKM elements.
$T$, $C$, $E$ and $A$ stand for color-allowed tree, color-suppressed tree,
$W$-exchange and annihilation topologies, respectively.
The primed symbols indicate the decays with a kaon as the bachelor particle in the final state.
The factor $\sqrt{2}$ originates from the isospin decomposition of the neutral pion,
$|\pi^0> = (u\bar{u}-d\bar{d})/\sqrt{2}$.
}\label{tab:decays}
\end{center}
\end{table}

At tree level, the $B^+\to \bar{D}^0\pi^+$ amplitude receives contributions from a
color-allowed ($T$) and color-suppressed topology ($C$), whereas the $B^+\to
{D}^0\pi^+$ amplitude proceeds predominantly through the color-suppressed topology
($C^{ub}$) and also via the annihilation topology,
as illustrated in Fig.~\ref{fig:top},
where the superscript $ub$ indicates that the decay proceeds through a
$b\to u$ transition. 

The method to estimate \rdpi with $B^0\to\bar{D}^0K^0$ decays is given in Sec.~\ref{sec:dk},
whereas the use of $B^0\to\bar{D}^0\pi^0$ decays is shown in Sec.~\ref{sec:dpi}.
The effect of the annihilation diagram is estimated in Sec.~\ref{sec:A}.

\section{Estimating \rdpi from branching fractions}
\label{sec:rdpi}
The expression for the branching fraction takes the following form:
\begin{equation}
\mbox{BR}(B \to Dh) = |A(B\to Dh)|^2\, \Phi^d_{Dh} \, \tau_{B}, \nonumber
\end{equation}
where $h$ is a pion or a kaon, $\Phi^d_{Dh}$ is a phase-space factor, $A(B\to Dh)$
is the total amplitude (containing the CKM elements, form factors and decay
constants) and $\tau_B$ is the lifetime of the $B$ meson.
Contributions from ``rescattering'' (like $B^+\to D^-\pi^+ \to \bar{D^0}\pi^0$)
are small, as shown within the framework of QCD factorization by Beneke et al.~\cite{Beneke:2000ry}.

\begin{table}[!b]
\begin{center}
\begin{tabular}{rlr}
\hline
$V_{CKM}$  &     &  Ref. \\
\hline
$|V_{ud}|=$     & $0.97425 \pm 0.00022$          & \cite{PDG2014}  \\
$|V_{us}|=$     & $0.2253 \pm 0.0008$            & \cite{PDG2014}  \\
$|V_{ub}|=$     & $(3.72\pm 0.16) \cdot 10^{-3}$ & \cite{Lattice:2015tia} \\
\hline
$|V_{cd}|=$     & $0.225\pm 0.008$               & \cite{PDG2014}  \\
$|V_{cs}|=$     & $0.986\pm 0.016$               & \cite{PDG2014}  \\
$|V_{cb}|=$     & $(41.1\pm 1.3)\cdot 10^{-3}$   & \cite{PDG2014}  \\
\hline
\end{tabular}
\caption{Values of CKM elements used.}\label{tab:Vckm}
\end{center}
\end{table}

The following estimate of the ratio of amplitudes can be made,
\begin{eqnarray}
\label{eq:r_TC}
\rdpi & = & \frac{A(B^+\to {D}^0\pi^+)}{A(B^+\to \bar{D}^0\pi^+)} = \frac{|C^{ub}|}{|T+C|}  \nonumber \\
      & = & \Big|\frac{V_{ub}V_{cd}}{V_{cb}V_{ud}}\Big| \frac{z \, |C|}{|T+C|}
\end{eqnarray}
where $z$ quantifies the ratio between the hadronic parts of the
two color-suppressed tree diagrams proceeding through a
$b\to c$ or $b\to u$ transition (shown in Fig.~\ref{fig:top}),
$C^{ub}=z\, C\times (V_{ub}V_{cd})/(V_{cb}V_{ud})$.
The contribution from the annihilation topology is also absorbed in the quantity $z$,
and will be further discussed in Sec.~\ref{sec:A}.

We can estimate $|C|/(|T|+|C|)$ in two ways,
\begin{itemize}
\item[A)] $\rdpi  \sim A(B^0\to \bar{D}^0 K^0)/A(B^+\to \bar{D}^0 K^+)$,
applying SU(3) symmetry, and correcting for the different CKM-elements involved;
\item[B)] $\rdpi \sim A(B^0\to \bar{D}^0 \pi^0)/A(B^+\to \bar{D}^0 \pi^+)$,
using external estimates for the contribution from $W$-exchange topologies ($E$)
to the decay $B^0\to \bar{D}^0 \pi^0$.
\end{itemize}

The magnitude of $z$ will be estimated in Sec.~\ref{sec:rDK} by comparing the result of the
amplitude ratio of the decays $B^+\to D^0K^+$,  \rdk, to the measured value by
LHCb~\cite{LHCb-PAPER-2016-032}.
For the numerical values of the CKM elements, we use the values listed in Table~\ref{tab:Vckm}.

\subsection{Estimating $\rdpi$ from \texorpdfstring{$B^0\to\bar{D}^0K^0$}{}}
\label{sec:dk}
The decays $B\to DK$ can be used to estimate the contributions of various $B\to D\pi$
decay topologies, assuming SU(3) symmetry.

The validity of this assumption was probed by comparing the $D^{(*)}K$ and $D^{(*)}\pi$ decay rates,
correcting for differences in phase space, CKM-elements, form factors and 
decay constants~\cite{Fleischer:2011}.
This assures that the decays $B^0\to \bar{D}^{0}K^0$ and $B^+\to \bar{D}^{0}K^+$
can be used to estimate a value for the amplitude ratio, $|C|/|T+C| = |C'|/|T'+C'|$, where
\begin{equation}
\label{eq:alpha}
\left| \frac{C'}{T'^{} +C'^{} } \right|
= \sqrt{\frac{\alpha \, \mathrm{BR}(B^0\to \bar{D}^{0}K^0)}{\mathrm{BR}(B^+\to \bar{D}^{0}K^+)}}.
\end{equation}
The factor $\alpha$ quantifies a correction to the quoted value of
$\mathrm{BR}(B^0\to \bar{D}^{0}K^0)$ in the PDG~\cite{PDG2014}. The measured branching fraction by
the BaBar~\cite{Aubert:2006qn} and Belle~\cite{Krokovny:2002ua} collaborations is obtained from the
sum over the charged-conjugate final states,
and therefore the quoted branching fraction represents the sum of the
$B^0\to\bar{D}^{0}K^0$ and $B^0\to {D}^{0}K^0$ branching fractions.
Recently LHCb also performed an analysis of the decays $B_{(s)}^0\to \bar{D}^{0}K^0$~\cite{LHCb-PAPER-2015-050}.

The quoted branching fraction can thus be expressed as the sum of the squares of the 
two color-suppressed tree amplitudes,
\begin{eqnarray}
\label{eq:ADK}
\mathrm{BR}(B^0\to \brabar{D}^{0}K^0) & = & A(B^0\to \bar{D}^0 K^0)^2 + A(B^0\to {D}^0 K^0)^2 \nonumber \\
                             & = &  |C'|^2 + |C'^{ub}|^2   \nonumber \\
                             & = &  \Big(1+ z' \Big| \frac{V_{ub}V_{cs}}{V_{cb}V_{us}}\Big|^2\Big) \times |C'|^2  \nonumber \\
                             & = &  (1 + 0.156\, z') \times |C'|^2 \, ,
\end{eqnarray}
where $z'$ quantifies the ratio between the hadronic parts of the two color-suppressed tree diagrams
proceeding through the $b\to u$ and $b\to c$ transitions,
$|C'^{ub}| =  z'\, C' \times |V_{ub}V_{cs}/V_{cb}V_{us}|$.
Hence, we need to correct the quoted branching fraction of the
decay $B^0\to \bar{D}^0 K^0$ to yield an estimate of the amplitude of $C'$, relative to $|T'+C'|$ with
$\alpha=1/(1+ 0.156\, z')$, to obtain
\begin{eqnarray}
\label{eq:rdpi1_form}
\rdpi  & = & \Big|\frac{V_{ub}V_{cd}}{V_{cb}V_{ud}}\Big|
         \frac{z'}{1+ 0.156\, z'}
         \sqrt{\frac{\mathrm{BR}(B^0\to \bar{D}^{0}K^0)}{\mathrm{BR}(B^+\to \bar{D}^{0}K^+)}}.
\end{eqnarray}

\subsection{Estimating $\rdpi$ from \texorpdfstring{$B^0\to \bar{D}^0 \pi^0$}{}}
\label{sec:dpi}
A second estimate of $\rdpi$ can be obtained using the decay $B^0\to \bar{D}^0 \pi^0$.
The decay $B^0\to \bar{D}^0 \pi^0$ receives contributions from the color-suppressed tree diagram ($C$)
and from the $W$-exchange diagram ($E$).
The comparison of the $B^0\to \bar{D}^0 \pi^0$ and $B^+\to \bar{D}^0 \pi^+$
decay rates gives~\cite{Fleischer:2011}:
\begin{equation}
\left| \frac{C-E}{T+C} \right|
= \sqrt{2}\sqrt{\frac{\mathrm{BR}(B^0\to \bar{D}^{0}\pi^0)}{\mathrm{BR}(B^+\to \bar{D}^{0}\pi^+)}}
= 0.331 \pm 0.010 (\mathrm{BR}) 
\end{equation}
again assuming that CKM elements, form factors, decay constants and phase space factors cancel in the ratio.
The uncertainty originates from the uncertainty on the measured branching fractions.
The factor $\sqrt{2}$ originates from the isospin decomposition of the neutral pion.
Although the branching fraction $\mathrm{BR}(B^0\to \bar{D}^{0}\pi^0)$ is determined as the sum of the
$D^0$ and $\bar{D}^{0}$ final states, the $b\to u$  color-suppressed tree amplitude is negligible compared to the
$b\to c$ amplitude, unlike the situation of Eq.~(\ref{eq:alpha}).

The color-suppressed tree diagram is expected to dominate the
total transition amplitude with respect to the $W$-exchange topology $E$,
which is supported by the comparison of the branching fractions of $B^0\to \bar{D}^0 \pi^0$
and $B^0\to \bar{D}^0 K^0$~\cite{Fleischer:2011}, leading to
\begin{equation}
\left|\frac{C-E}{C}            \right|   =  0.913 \pm 0.074.
\label{eq:C-E}
\end{equation}
To obtain an independent estimate of \rdpi with respect to Eq.~(\ref{eq:rdpi1_form}),
{\em i.e.} without re-using information on the branching fraction of $B^0\to \bar{D}^0 K^0$,
the size of the $W$-exchange amplitude can be estimated from the decay
$B^0\to {D}_s^- K^+$~\cite{LHCB-PAPER-2014-064}, resulting in the following value~\cite{Fleischer:2011},
\begin{equation}
\left|\frac{E}{T+C}\right| =  0.056 \pm 0.004.
\label{eq:E-TC}
\end{equation}
Without assuming any value for the relative phase between the $W$-exchange ($E$) and
color-suppressed ($C$) amplitudes, we assign the full contribution of the $W$-exchange
amplitude as uncertainty to the estimate of $|C/|T+C|$,
\begin{equation}
\left|\frac{C}{T+C}\right| = 0.331 \pm 0.010 (\mathrm{BR}) \pm 0.056 (E).
\end{equation}
The resulting expression for $\rdpi$ then becomes
\begin{eqnarray}
\label{eq:rdpi2_form}
\rdpi  & = & \Big|\frac{V_{ub}V_{cd}}{V_{cb}V_{ud}}\Big|
         \sqrt{2}\, z \, \sqrt{\frac{\mathrm{BR}(B^0\to \bar{D}^{0}\pi^0)}{\mathrm{BR}(B^+\to \bar{D}^{0}\pi^+)}}.
\label{eq:rdpi-2}
\end{eqnarray}

\subsection{Effect of annihilation topology}
\label{sec:A}
The relative contribution from the annihilation topology with respect to the 
color-suppressed tree topology for the $B^+\to {D}^0\pi^+$ amplitude,
is estimated using the measured branching fraction of the decay $B^+\to D_s^+\phi$~\cite{LHCb-PAPER-2012-025},
relative to the decay $B^0\to D^0K^0$.
At lowest order the $B^+\to D_s^+\phi$ decay  proceeds purely through the annihilation topology.
 
The estimate of $|A/C|$ for the $B^+\to {D}^0\pi^+$ case can be directly obtained from the branching ratios, 
when corrected for by the appropriate CKM-elements and decay constants $f_X$,
\begin{eqnarray}
|A/C|   & = & \sqrt{\frac{\mathrm{BR}(B^+\to D_s^+\phi)}{\mathrm{BR}(B^0\to D^0K^0)}}  
              \Big( \frac{V_{cb}V_{us}}{V_{ub}V_{cs}} \Big)
              \Big( \frac{f_D f_K}{f_{Ds} f_\phi} \Big)  \sim  0.25 \nonumber
\end{eqnarray}
with a large uncertainty from the branching fraction measurement of $B^+\to D_s^+\phi$,
see Tab.~\ref{tab:decays}. It is also noted that the branching fraction 
$\mathrm{BR}(B^+ \to \bar{D}^0D_s^+)$ deviates from $\mathrm{BR}(B^0 \to D^-D_s^+)$, where the main
difference is expected to arise from the annihilation contribution~\cite{Bel:2015wha}.
Possible contributions to these final states from rescattering processes are 
discussed in Ref.~\cite{Gronau:2012gs}.
The relative phase between the annihilation and color-suppressed tree topology is unknown, so 
the annhilation contribution can enhance or reduce the value of \rdpi.
Assuming SU(3) symmetry, this contribution is expected to be equal in the 
$B^+\to {D}^0K^+$ system, and will thus be accounted for in the 
determination of $z$ from \rdk in the next Section.

\section{Correction using \rdk}
\label{sec:rDK}
To quantify the ratio $z$ between the hadronic parts of the $b\to u$ and $b\to c$ color-suppressed
tree diagrams,
$C'^{ub}=z'\, C' \times (V_{ub}V_{cs}/V_{cb}V_{us})$, an estimate for \rdk can be obtained in a similar way,
and be compared to the fitted value for \rdk from the LHCb fit~\cite{LHCb-PAPER-2016-032}.
The quantity $z$ also contains the correction due to contributions from the annihilation topology,
see Fig.~\ref{fig:top}.
We obtain the following expression for \rdk,
\begin{eqnarray}
\label{eq:rDK_TC}
\rdk  & = & \Big|\frac{V_{ub}V_{cs}}{V_{cb}V_{us}}\Big|
         \frac{z'}{1+ 0.156\, z'}
         \sqrt{\frac{\mathrm{BR}(B^0\to \bar{D}^{0}K^0)}{\mathrm{BR}(B^+\to \bar{D}^{0}K^+)}} ,  \nonumber
\end{eqnarray}
which differs from Eq.~(\ref{eq:rdpi1_form}) by different CKM elements involved.
Inserting the value for \rdk obtained from the LHCb fit, $\rdk = 0.101 \pm 0.006$~\cite{LHCb-PAPER-2016-032},
the following estimate for the ratio of the hadronic parts of the color-suppressed amplitudes is obtained,
\begin{eqnarray}
\frac{z'}{1+ 0.156\, z'}  =  0.68 \pm 0.05 & \Rightarrow & z' =  0.76\pm 0.07.
\end{eqnarray}
The fact that the value of $z'$ is close to unity, indicates that the hadronic parts of
the two color-suppressed tree diagrams are of similar magnitude, in particular 
if the annihilation topology negatively interfers with the color-suppressed tree topology,
i.e. if the relative strong phase is close to 180$^o$, which would lead to a value $z'\sim 0.75$.
We assume that the deviation from unity is equal for the $D\pi$ case, with an uncertainty of
10\% from SU(3) symmetry breaking effects, $z = 0.76 \pm 0.07(\mathrm{BR}) \pm 0.02(\mathrm{SU(3)})$.

Inserting the numerical values in Eq.~(\ref{eq:rdpi1_form}) and Eq.~(\ref{eq:rdpi2_form}) leads to the following
estimates of \rdpi,
\begin{eqnarray}
\rdpi(D^0K^0) & = & 0.0053 \pm 0.0002(\mathrm{V_{CKM}}) \pm 0.0004(\mathrm{BR}) \nonumber   \\
      &   & \hspace{1.2cm}       \pm \,  0.0005(\mathrm{SU(3)}) \label{eq:rdpi1}\\
\rdpi(D^0\pi^0) & = & 0.0053 \pm 0.0002(\mathrm{V_{CKM}}) \pm 0.0002(\mathrm{BR}) \nonumber  \\
      &   & \hspace{1.2cm}       \pm \,  0.0009(\mathrm{E})    \pm \,  0.0005(z)
\end{eqnarray}
which are in good agreement,
albeit with a large uncertainty from the $W$-exchange contribution to the $B^0\to \bar{D}^0 \pi^0$
decay rate.
The agreement shows the internal consistency of the approach presented here.
An additional 10\% uncertainty from SU(3) symmetry is assumed in the estimate
of Eq.~(\ref{eq:rdpi1}), based on the agreement
of the relative contributions of the various decay topologies to the
$B\to DK$ and $B\to D\pi$ decays~\cite{Fleischer:2011}.
Given the correlated systematic uncertainties between the two results, the following
combined estimate is obtained,
\begin{equation}
\rdpi = 0.0053 \pm 0.0007. \nonumber
\end{equation}
\vspace{-1.0cm}

\section{Conclusions}
The estimate for the value of the amplitude ratio \rdpi that is presented here
provides a valuable input to the discussion of the measurement of \rdpi at
LHCb. The actual measurement of \rdpi can be achieved either by a combination of
indirect measurements, as presented in
Refs.~\cite{LHCb-CONF-2014-004,LHCb-PAPER-2016-032}, or by direct measurement
using semileptonic decays of the form $B^{+}\to D^{0}\pi^{+}$, where $D^{0}\to
K^{-}\mu^{+}\nu_{\mu}$, and the charge of the kaon and muon can unambiguously
tag the $D^0$ flavor.  Future determinations of \rdpi can be compared to the
estimate presented here, to assess the validity of the assumptions on
rescattering and SU(3) symmetry as used in this paper.  The LHCb collaboration
foresees to accumulate a four times larger dataset by the end of 2018, and a
five time smaller uncertainty at the end of the LHCb upgrade, which will result
in an experimental  uncertainty of the measured value of \rdpi that is  smaller
than the one presented here.

\noindent{\bf Acknowledgments} \\
We would like to thank Robert Fleischer, Vincenzo Vagnoni and Greg Ciezarek 
for many valuable discussions.

\end{document}